\documentclass[article, prX]{revtex4}
\usepackage{graphicx}
\begin{document}
\title{Measuring Light Pollution with Fisheye Lens Imagery from A Moving Boat, A Proof of Concept}

\author{Andreas Jechow}
\affiliation{Ecohydrology, Leibniz Institute of Freshwater Ecology and Inland Fisheries, M{\"u}ggelseedamm 310, 12587 Berliny}
\affiliation{Remote Sensing, Helmholtz Center Potsdam, German Center for Geosciences GFZ, Telegraphenberg, Potsdam}
\author{Zolt{\'a}n Koll{\'a}th}
\affiliation{Eötvös Loránd University, Savaria Department of Physics, Károlyi Gáspár tér 4, 9700 Szombathely, Hungary}
\author{Amit Lerner}
\affiliation{National Institute of Oceanography, Israel Oceanographic and Limnological Research, Haifa, 31080, Israel}
\author{Franz H{\"o}lker}
\affiliation{Ecohydrology, Leibniz Institute of Freshwater Ecology and Inland Fisheries, M{\"u}ggelseedamm 310, 12587 Berliny}
\author{Andreas H{\"a}nel}
\affiliation{Museum am Sch{\"o}lerberg, Klaus-Strick-Weg 10, Osnabrück, 49082, Germany}
\author{Nadav Shashar}
\affiliation{Ben-Gurion University of the Negev, Eilat Campus, Beer-Sheva, 84105, Israel}
\author{Christopher C. M. Kyba}
\affiliation{Remote Sensing, Helmholtz Center Potsdam, German Center for Geosciences GFZ, Telegraphenberg, Potsdam}
\affiliation{Ecohydrology, Leibniz Institute of Freshwater Ecology and Inland Fisheries, M{\"u}ggelseedamm 310, 12587 Berliny}

%

\maketitle

\section*{Abstract}
Near all-sky imaging photometry was performed from a boat on the Gulf of Aqaba to measure the night sky brightness in a coastal environment. The boat was not anchored, and therefore drifted and rocked. The camera was mounted on a tripod without any inertia/motion stabilization. A commercial digital single lens reflex (DSLR) camera and fisheye lens were used with ISO setting of 6400, with the exposure time varied between 0.5 s and 5 s. We find that despite movement of the vessel the measurements produce quantitatively comparable results apart from saturation effects. We discuss the potential and limitations of this method for mapping light pollution in marine and freshwater systems. This work represents the proof of concept that all-sky photometry with a commercial DSLR camera is a viable tool to determine light pollution in an ecological context from a moving boat.

\section{Introduction}
Artificial light at night (ALAN) allows humans to extend activities up to a 24 hour-a-day period. While indoor use of ALAN can affect human health as it disrupts the natural circadian rhythm and suppresses melatonin production \cite{Stevens:2015}, the direct photo-related environmental consequences are usually small as the light is mainly confined in buildings. Outdoor lighting on the other hand may have less dramatic consequences on human health \cite{Stevens:2015}, but ALAN can spill into the naturally nocturnal landscapes and cause light pollution (LP). The increase of outdoor ALAN is a global phenomenon with growth rates of 3-6 $\%$ per year in industrialized countries and higher in developing regions \cite{book:Narisada, Hoelker:2010_b}. Sustainable lighting technology may help to reduce the overall LP \cite{Lyyti:2015}. Several studies have investigated the impact of ALAN on the environment \cite{book:rich_longcore} and there is a growing concern that LP affects biodiversity \cite{Hoelker:2010_a, Gaston:2015_ptb}. Until recently, the main focus of studies about ecological LP was on terrestrial animals \cite{book:rich_longcore, rowse2016dark, robert2015artificial, macgregor2015pollination} and plants \cite{macgregor2015pollination, bennie2016plants}. However, because human settlements concentrate along freshwater reservoirs and coastlines, research on ecological LP has shifted towards aquatic systems \cite{davies2014marine, Perkin:2011, Moore:2000, bruening2015spotlight, Jechow2016}.

Historically, the first evaluations of the night sky brightness (NSB) in the context of LP were performed by astronomers \cite{Riegel1973, berry1976light}. They concentrated their investigations on clear sky conditions and the ability to observe celestial objects. However, illumination conditions due to LP and artificial skyglow (the backscattered portion of the upwelling ALAN) vary dramatically with meteorological conditions, especially clouds \cite{Kyba:2011_sqm, Kocifaj:2014_cloud_theory, jechowALAN,  Ribas2016clouds}. Recent work on clouds was linking NSB data with laser ceilometer measurements \cite{Ribas2016clouds}. The combination of satellite data and radiative transfer modelling can provide accurate estimates of the NSB for clear skies \cite{falchi2016WA}, but is not (yet) applicable to overcast situations. Ground based NSB measuring methods can provide information in overcast conditions. These include single sensors such as inexpensive handheld sky brightness meters \cite{Cinzano_sqm:2005} with the advantage of providing data from citizen scientists, hobby astronomers and researchers on a local to global scale \cite{Jechow2016, Kyba:2015_isqm, Pun:2014}. The drawback of this method is that no spatial and no spectral information is available, and most devices only measure the NSB at zenith, potentially missing out higher fractions of LP near the horizon \cite{Jechow2016, jechowALAN, kocifaj2015zenith}. Photometry with DSLR cameras and fisheye-lenses \cite{Jechow2016, Kollath:2010} or mosaics \cite{Duriscoe:2007} can provide this spatial information about the NSB, and have become available for the public at reasonable prices since consumer electronics has become a mass phenomenon.

Usually all-sky photometry as part of astro-photometry is done in a terrestrial context using special filters, long exposure times and fixed tripods or mounts, potentially involving even compensation for the Earth’s rotation \cite{Duriscoe:2007}. In astro-photometry, well known objects are used as references for extinction measurements and the studies very often concern single stars and very dark sites \cite{romanishin2006introduction}. A very interesting and powerful device in this context is the ASTMON (All-Sky Transmission MONitor) making use of all-sky imaging in several spectral bands with a filter wheel \cite{aceituno2011all}, which was also used in areas with LP \cite{aube2016spectral}. For the measurement of the NSB in the context of ecological LP and to compare different sites this precision is not necessary. The effects of ALAN can vary by orders of magnitudes within a radius of several tens of kilometers \cite{jechowALAN, Pun:2014} or by weather conditions \cite{Jechow2016, Ribas2016clouds}. For most species, the amount of light is the important parameter rather than single celestial objects. However, visual animals might react on the directionality that cannot be observed by single sensor devices like lux meters.
NSB measurements with either single sensors or all-sky photometry are only sparsely done in an aquatic or even in a marine context \cite{Kyba:2015_isqm}. We have recently quantified the NSB at a freshwater lake in Germany from a floating platform using both methods \cite{Jechow2016}. However, the platform was anchored and rocking only slightly. Here we demonstrate that (near) all-sky photometry can also be done from a moving boat in a marine context by using a commercial DSLR camera with a fisheye lens. The camera was mounted on a tripod, without using motion-compensation. Despite this, we obtained NSB measurements with spatial information from almost the whole hemisphere. In this proof-of-concept study, we discuss the limitations and the potential of this method in the context of applying it to investigate the spread of LP, and in particular skyglow, from coastal towns into unpolluted areas across open waters.

\section{Methods}
\subsection{The study site}
The measurements were performed during astronomical night on the 5th of March 2016 near the city of Eilat in Israel on the Gulf of Aqaba close to the Interuniversity Institute for Marine Sciences (IUI) at 29$^{\circ}$29'33.0"N, 34$^{\circ}$55'54.8"E (see Fig. 1) from an 6.7 m long and 3.45 m wide single engine boat. Parts of the western shoreline north of the IUI are a marine protected area: Eilat's Coral Beach Nature Reserve and Conservation area, a nature reserve and national park in the Red Sea. It covers 1.2 kilometers of shore, and is the northernmost coral reef in the world.

\begin{figure}[h]
\centering
\includegraphics[width=160mm]{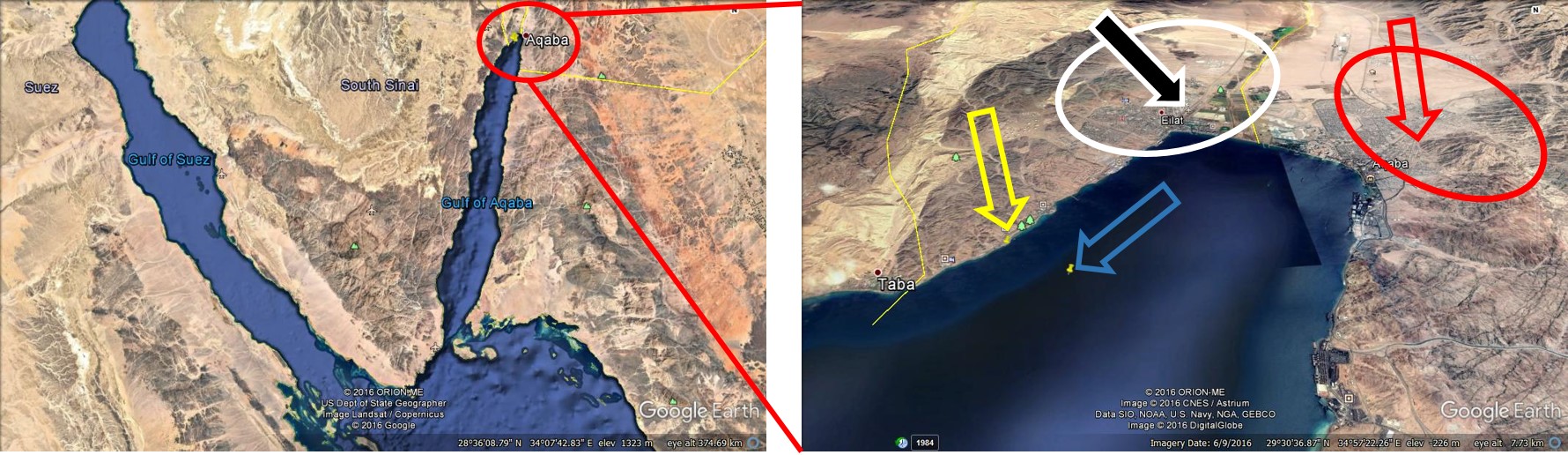}
\caption{NSB measurement site. Left image: the Gulf of Aqaba with the Sinai Peninsula on the left and Jordan and Saudi Arabia on the right. Red circle marks the measurement region. Right image – the northern tip of the Gulf of Aqaba where the fisheye lens observations were made. The blue arrow represents the measurement location near the border of Egypt, Jordan and Israel. The red arrow and circle indicate the city of Aqaba, the white arrow and circle indicate the city of Eilat and the yellow arrow shows the position of the Interuniversity Institute for Marine Science (IUI).}
\label{site}
\end{figure}

\subsection{Digital single lens reflex camera with fisheye lens}
NSB measurements can be performed with calibrated DSLR cameras that allow saving images in an unaltered raw format \cite{Kollath:2010, KybaLonne2015, RibasLonne2017}. After calibration, it is possible to convert the camera’s radiance observations in digital numbers to photopic luminance, and to produce false color images that show spatially resolved NSB in units of cd/m$^2$, mag$_{SQM}$/arcsec$^2$ or natural sky units (NSU), respectively. If carefully calibrated, it is possible to reach 10 percent precision with a commercial DSLR camera \cite{Kollath:2010}. The NSU is a useful translation of NSB into a relative unit, comparing how much a site differs from a relatively unpolluted site defined to have a value of 1 NSU. It was introduced to allow non-astronomers to quickly rate the amount of LP at different locations (see e.g. \cite{Kyba:2015_isqm}) using values from the widely used sky quality meter (SQM, Unihedron, Canada). NSU =  1 is defined at 21.6 mag$_{SQM}$/arcsec$^2$, and 254µcd/m$^2$, respectively. It can be transformed by using the equation: NSU=(10)$^{0.4(21.6-X)}$, where X is the NSB in mag$_{SQM}$/arcsec$^2$. Please note that the NSU is not arbitrary but based on a real photometric quantity “magnitudes” dating back to Pogsons recommendation in 1856 \cite{pogson1856magnitudes} and that there is a deviation between Johnson V-band (used for magnitudes in astronomy), the photopic response curve of the human eye, the response curve of the SQM and the DSLR cameras spectral response \cite{haenel2017}. 
We used a Canon EOS 6D camera with a Sigma EX DG circular fish-eye lens. The lens has a focal length of 8 mm and an aperture of F3.5. The camera has a 20.2 Megapixel full frame (36 mm x 24 mm) CMOS sensor and an integrated GPS tracker. The camera was cross-calibrated with a thoroughly calibrated camera during an earlier measurement campaign \cite{KybaLonne2015} and several DSLR cameras in a recent intercomparison campaign \cite{RibasLonne2017}. Pictures were obtained in full format (5472 pixel x 3648 pixel), with ISO setting of 6400, and a varied exposure time between 0.5 seconds and 5 seconds. All images were saved in raw format. The images were processed using ``DiCaLum Ver. 0.9'' \cite{kollath2017} an open source code based on the free software GNU Octave \cite{eaton1997gnu}.

\section{Results and discussion}
\subsection{Fisheye-lens images}
In Figure 2, (near all-sky) fisheye-lens pictures taken from the slightly rocking and turning boat are shown for different exposure times a) 5 seconds (image $\#$1517), b) 2 seconds (image $\#$1518), c) 1 second (image $\#$1519) and d) 0.5 seconds (image $\#$1520). All four pictures were taken at 21:42 pm local time (19:42 pm GMT). The camera was orientated approximately towards the northeast/southwest axis of the Gulf, with the lower part of the image pointing to the northeast. The camera was not aimed directly towards zenith in order to illustrate the light reflected from the water surface. Therefore, it is not exactly all-sky imaging. The two distinct light domes that are visible at the bottom in Fig 2 a) are from the center of Eilat (white arrow with black filling on the lower left) and Aqaba (red arrow without filling on the lower right).
The rocking of the boat was relatively strong and is apparent by a distinct ``smearin'' of the stars in the pictures. To illustrate the possible impact of this rocking on the NSB measurements, Fig. 3 a)-d) shows zoom in pictures from the same images as in Fig 2 a)-d). The upper left area of the images was chosen, which is pointing to the southwest and included Sirius (magnitude -1.46, marked with a red circle) and the constellation of Orion (indicated with a light blue circle).
The constellation Orion is best visible in Fig. 3 c) and d) with low smearing. Here, the belt of three stars (Alnitak, Alnilam and Mintaka) aligned on an axis with Sirius can be seen. In all four images, the smearing can be tracked best with Sirius. In Fig. 3 a) and b), the smearing is strong and spiral, while in Fig. 3 c) the smearing is less strong and linear. In Fig. 3 d), the smearing is lowest but still perceptible by a linear distortion of the stars that should be point sources.
As expected, the longer the exposure time, the more susceptible to smearing by rocking and drifting. From the movement of the constellation between the different images, it can also be seen that the boat was slowly rotating clockwise. The zoomed in sections also show that the signal to noise ratio was slightly decreased when the exposure times were reduced. However due to the smearing the signal to noise improvement by increased exposure time is not as pronounced as with still images, as the signal now spreads over several different pixels instead of summing up on one pixel.
\begin{figure}[htbp!]
\centering
\includegraphics[width=143mm]{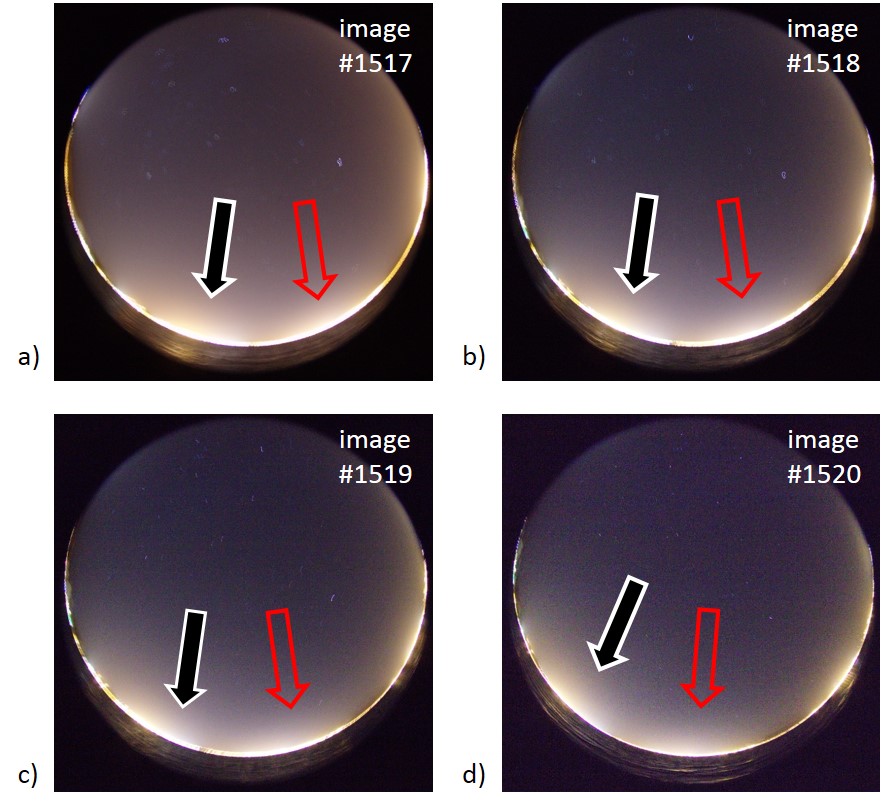}
\caption{Near all-sky pictures of the night sky above the study site obtained from a non-anchored, rocking boat using a commercial DSLR camera (Canon EOS 6D) with a fisheye lens (Sigma EX DG 8 mm). An ISO setting of 6400 was used with different exposure times of a) image $\#$1517 with 5 seconds exposure b) image $\#$1518 with 2 seconds exposure time c) image $\#$1519 with 1 second exposure time and d) image $\#$1520 with 0.5 seconds exposure time. Black and white arrow points to Eilat and red arrow to Aqaba.}
\vspace{2mm}
\centering
\includegraphics[width=143mm]{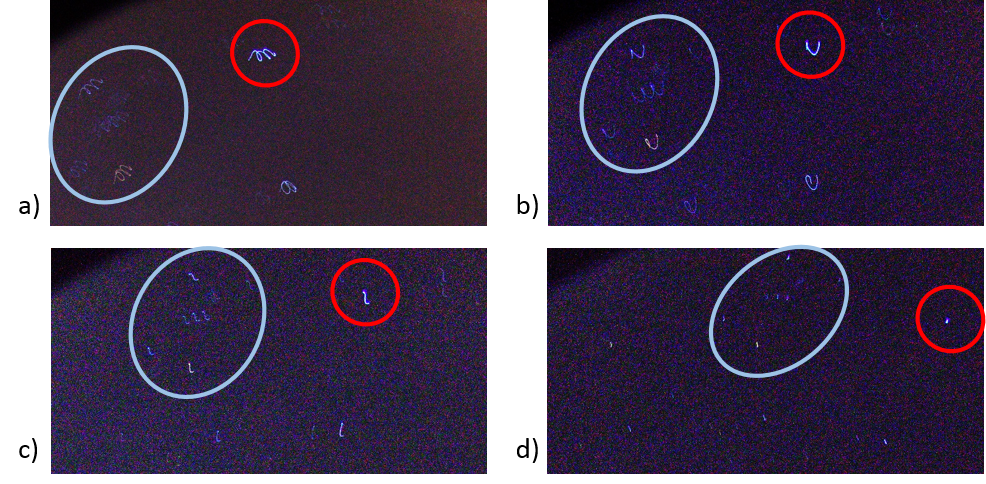}
\caption{Zoom in of the pictures of Figure 2 showing the detail of constellation of Orion (blue circle) and Sirius (red circle).}
\label{zoom}
\end{figure}

\subsection{Luminance maps}
The fisheye images have been converted to luminance maps that are shown in Figure 4 a)-d) on a scale ranging from 0.5 to 1000 NSU. The four luminance maps look relatively similar to each other. The zenith brightness was on the order of about 5 NSU and increased towards the horizon with peak brightness values exceeding 1000 NSU at the horizon, which is attributed to direct city lights. The light domes above the cities near the horizon reach NSB values of several hundred NSU. These large values indicate the night sky in this region was affected by skyglow.

Despite the qualitative similarities, there are some obvious differences in the luminance maps: First, the exposure time resulted in different distortion of the images due to the smearing as discussed above, and only slightly better signal to noise ratio for longer exposure times. The smearing leads to an “analog binning” effect that in conjunction with the better signal to noise ratio makes Fig. 4 a) appear most homogeneously or smoother than the other luminance maps. Fig. 4 d) is less smeared and has a lower signal to noise ratio, and therefore appears grainier. Second, for longer exposure times some pixels were overexposed, leading to saturation and loss of information about certain particularly bright regions near the horizon. This is most notable when comparing the light emission from the two cities in Fig. 4 a) and b). The high values in the range of 1000 NSU are resolved in b) but not in a). Third, the rotation of the boat resulted in a slightly different pointing for the luminance maps. This is most notable when comparing the brightness distribution at the water surface in the lower part of the luminance maps.

\begin{figure}[h]
\centering
a)\includegraphics[width=80mm]{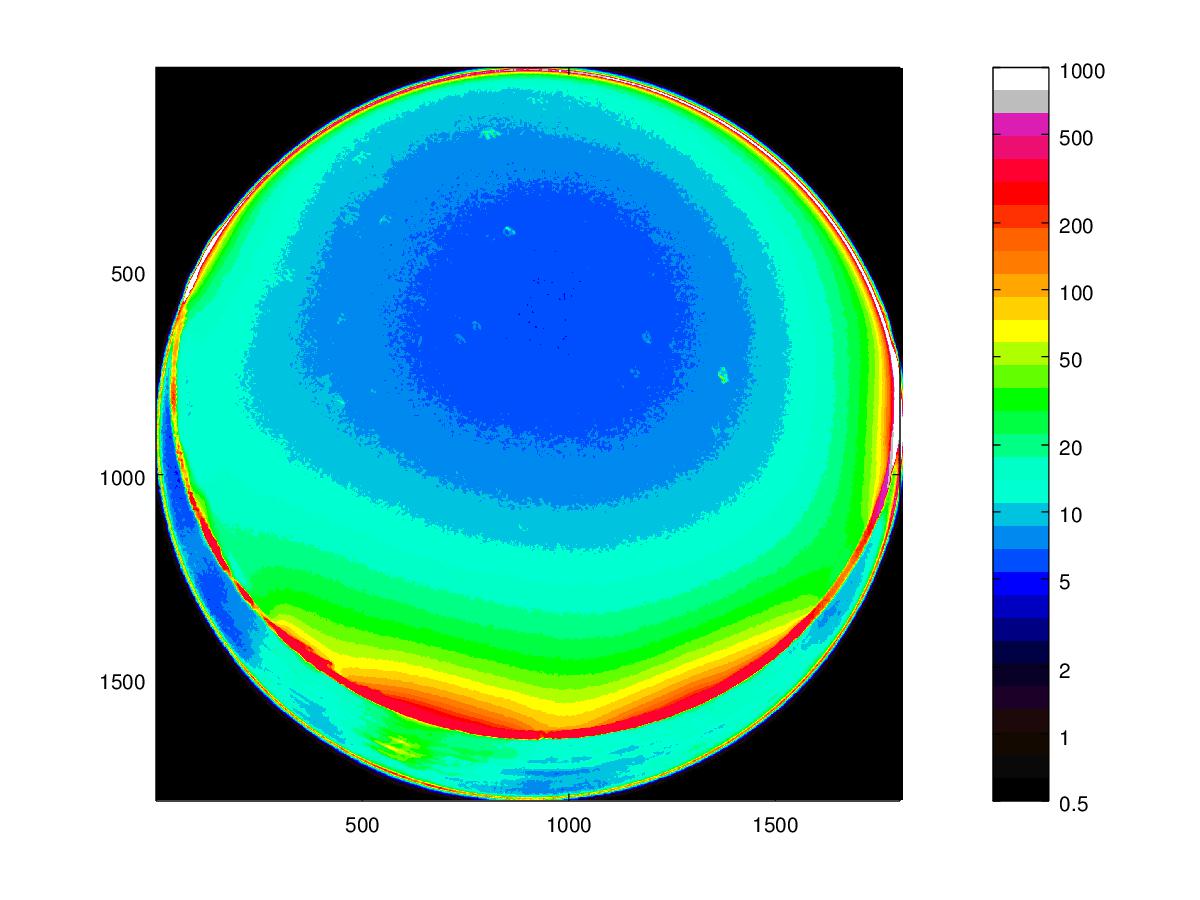}
b)\includegraphics[width=80mm]{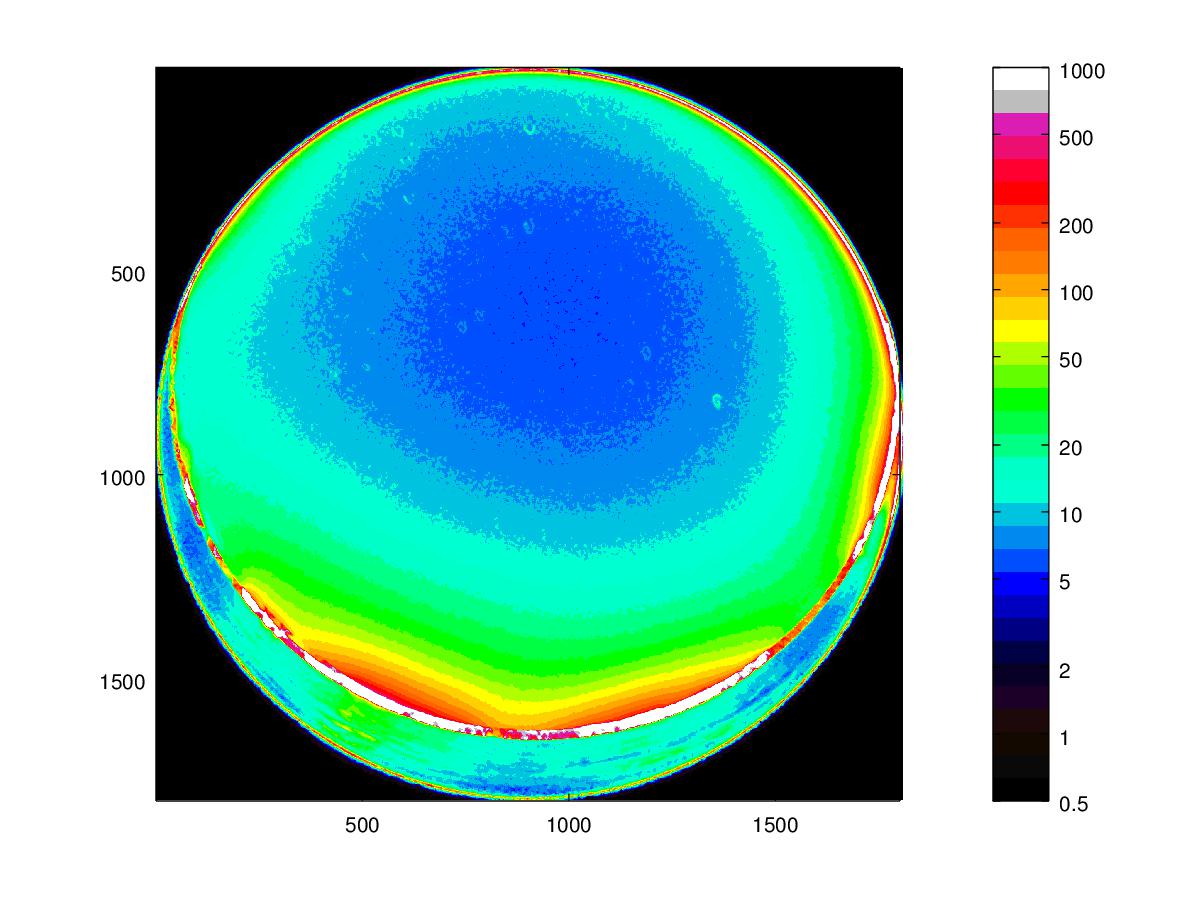}\\
c)\includegraphics[width=80mm]{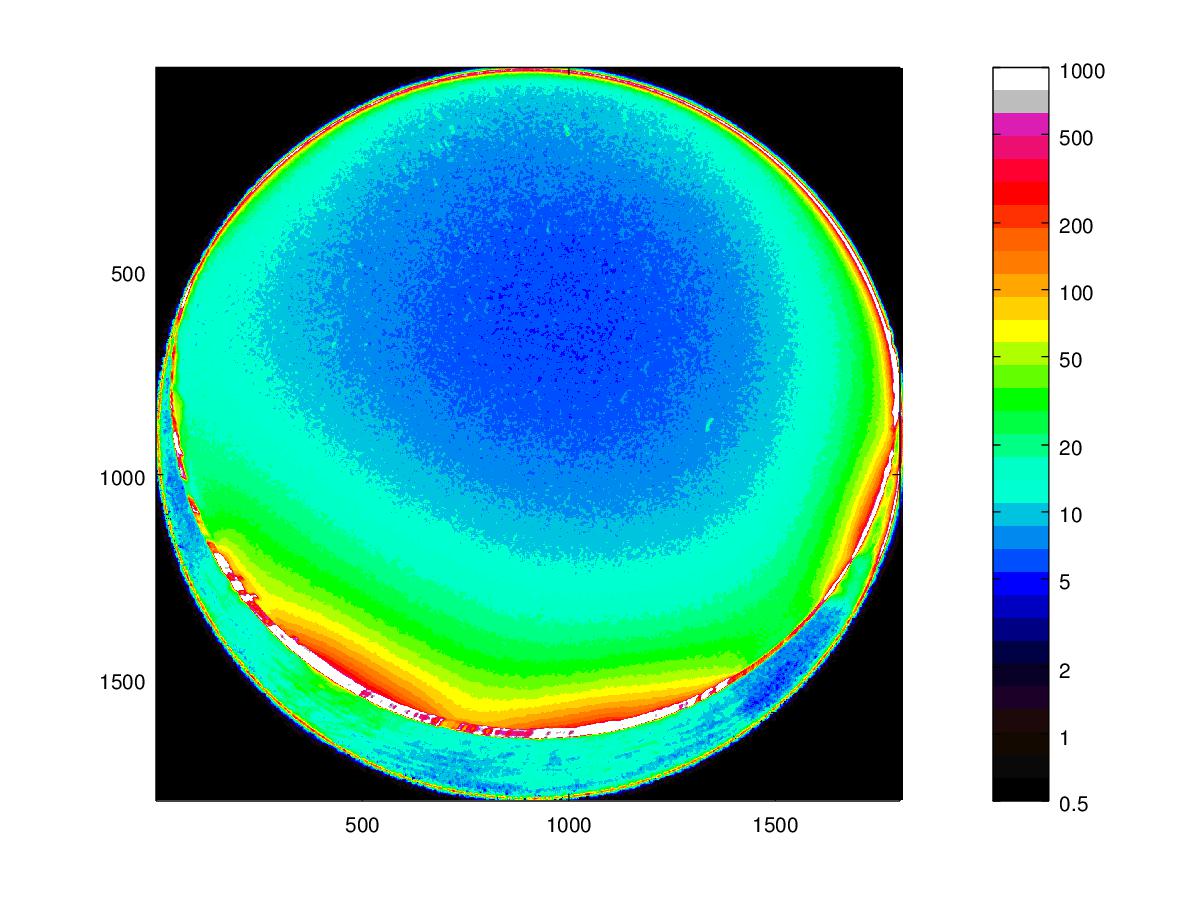}
d)\includegraphics[width=80mm]{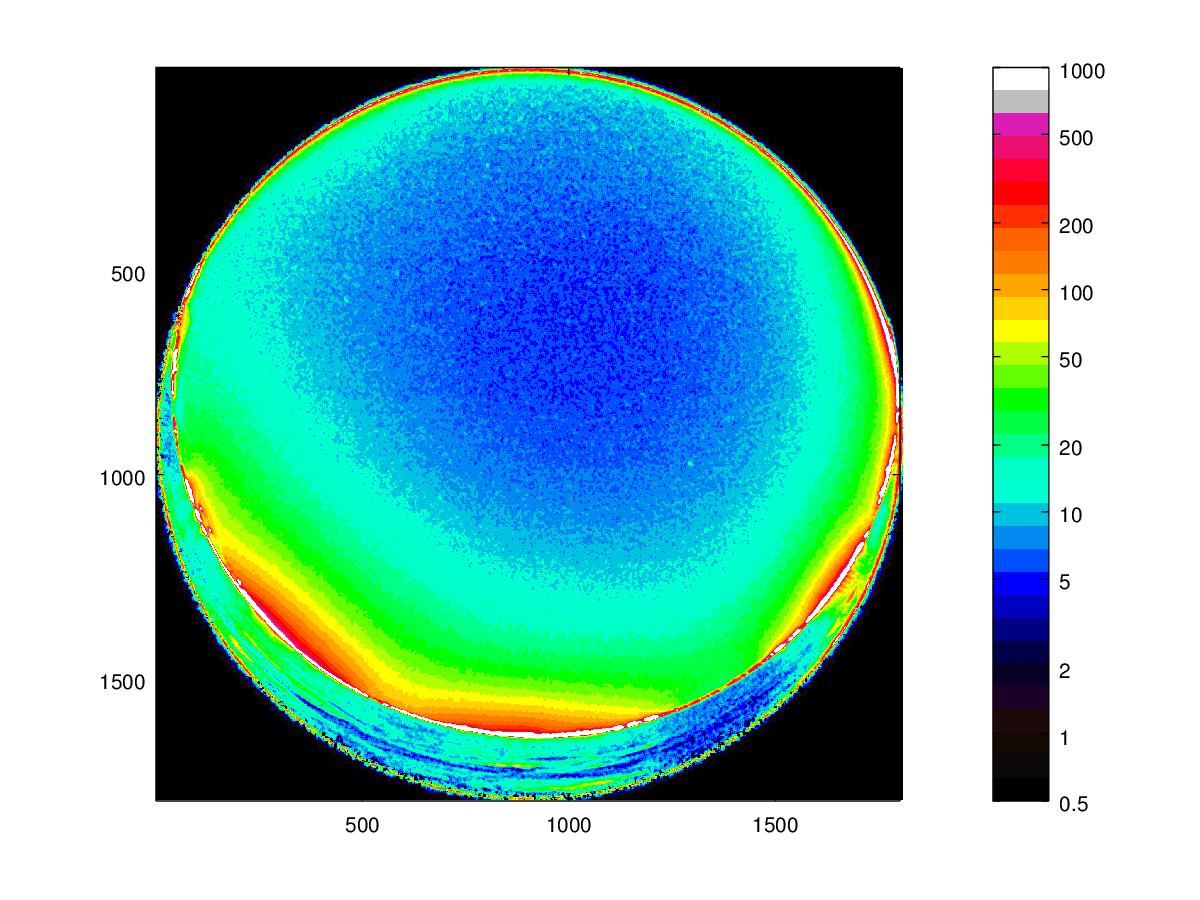}
\caption{Luminance maps calculated from the fisheye images of Fig. 2. The color bar represents NSU values. Note that NSU = 1 represents approximately a natural sky without light pollution (0.25 mcd/m$^2$).}
\label{images}
\end{figure}

\subsection{Angular luminance distribution}
As can be seen from the luminance maps, there is a brightness gradient from the zenith towards the horizon. To understand the limits of the method, it is useful to plot the angular brightness distribution. Figure 5 shows the luminance as a function of the angle with respect to the normal vector of the imaging plane (which is not exactly the zenith in our case, but for typical all-sky images normal vector should point to the zenith) for a) the full angular range and b) values up to 70$^{\circ}$. To obtain the angular luminance distribution, the brightness values for a specific angle have been averaged by integrating over the azimuth with DiCaLum \cite{kollath2017} using a 2$^{\circ}$. resolution in altitude (averaging over rings around the imaging normal vector, or zenith in true all-sky photometry).

The average brightness near the zenith is in the range of 6.0-6.4 NSU and increases to values of several tens of NSU towards the horizon. For angles up to 74$^{\circ}$, the luminance distribution is very similar between the measurements with different exposure times. Differences become apparent at higher angles, where saturation effects result in plateauing of the luminance values for the three longer exposure times due to overexposure. At overly long exposure times, the pixels saturate and therefore the average luminance is underestimated. With a static all-sky camera, this can be avoided by using the high dynamic range method (combining images with different exposure times). From a drifting and rocking boat, this is more complicated, but could be achieved by using the brightest stars for image co-registration. We judge that the short exposure time produces very good results, at least for this LP situation in the range of 5 NSU.
\begin{figure}[h]
\centering
a)\includegraphics[width=80mm]{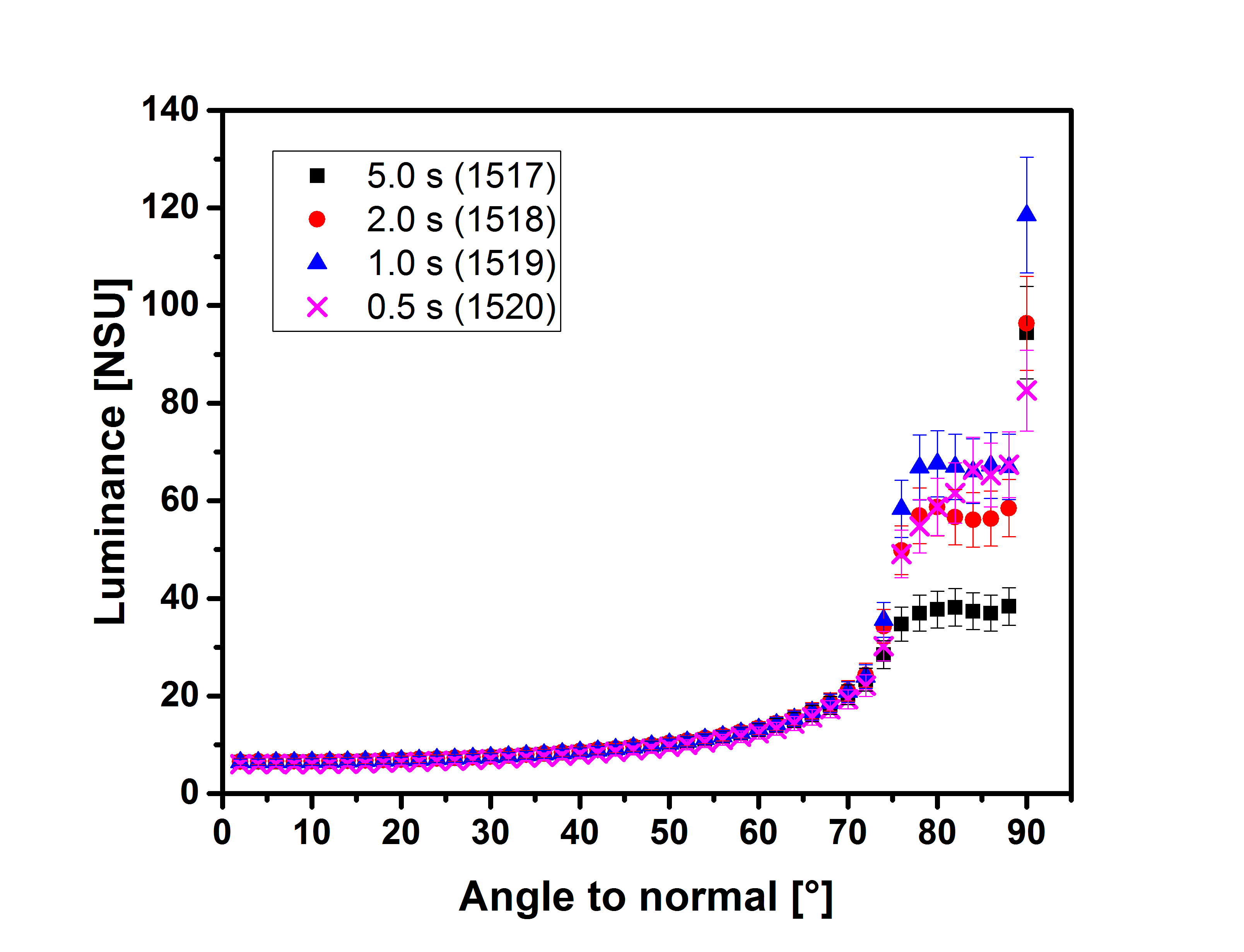}
b)\includegraphics[width=80mm]{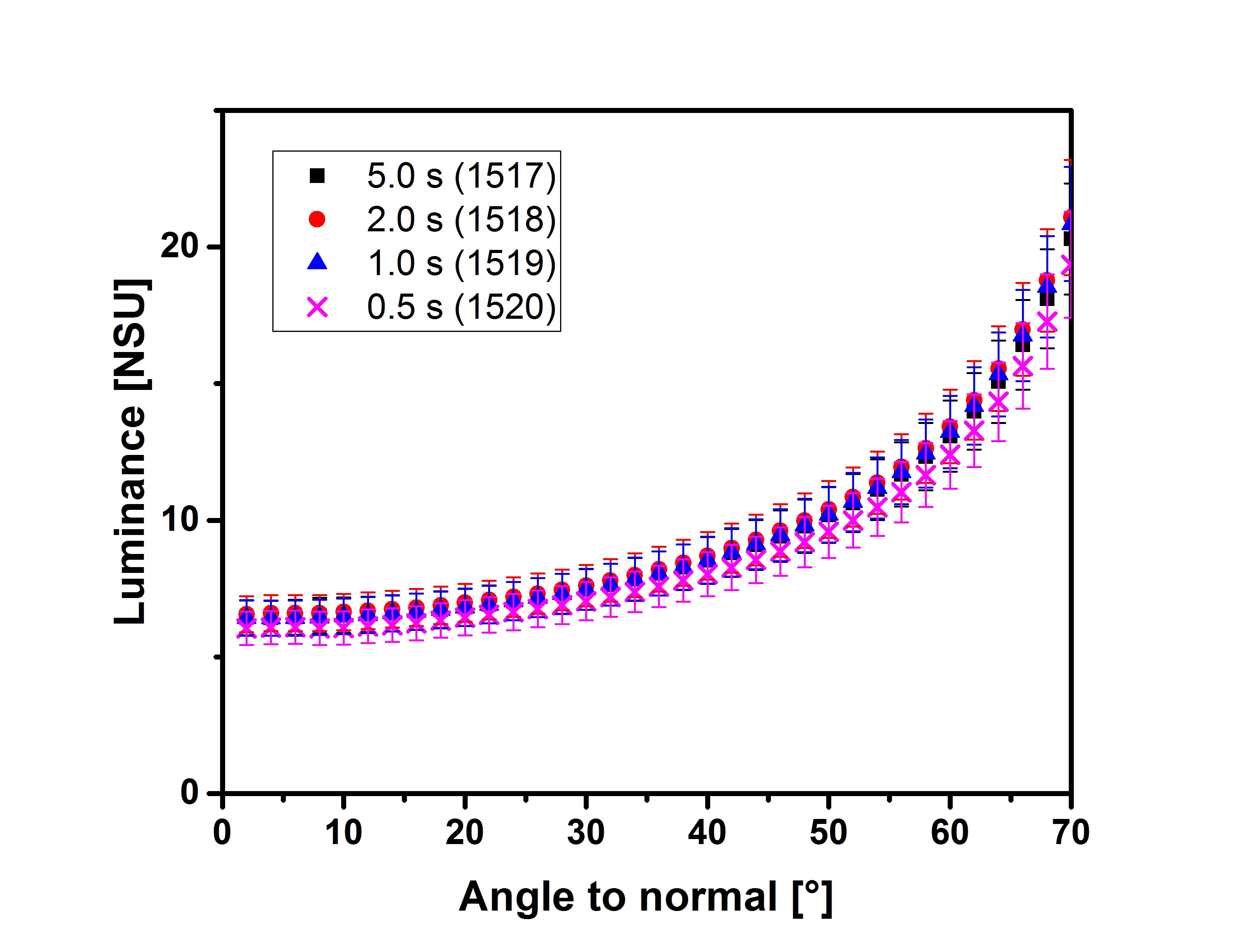}
\caption{Luminance as a function of the angle with respect to the imaging normal vector for, a) the full angular and dynamic range b) zoom in to angles below 70$^{\circ}$,where the values for different exposure times show very good agreement.}
\label{anglu}
\end{figure}

\subsection{Horizontal and scalar illuminance}
A common measure for the light incident at a specific site is the horizontal illuminance $E_{V,hor}$, usually obtained with lux meters. Assuming a homogeneous sky brightness, $E_{V,hor}$ can be inferred from NSB values in units of cd/m$^2$ (homogeneous luminance $L_{V,hom}$) by multiplying with $\pi$ \cite{kocifaj2015zenith}:

\begin{equation}
E_{V,hor} = L_{V,hom} \cdot \pi.
\end{equation}

However, this is not valid for most scenarios, as usually the NSB is not uniform \cite{duriscoe2016photometric, Jechow2016, jechowALAN, Kollath:2010}. Therefore, the horizontal illuminance can only be roughly approximated from narrow angle zenith measurements alone \cite{kocifaj2015zenith, duriscoe2016photometric}. From all-sky brightness maps, the horizontal illuminance (incident at the imaging plane of the camera) can be obtained when incorporating a cosine correction \cite{duriscoe2016photometric}. The cosine corrected angular luminance distribution is shown in Fig. 6 a) in NSU. The contribution of the NSB at the horizon is reduced and this results in a reduction of the impact of saturation effects for this particular LP situation. The cosine corrected NSB yields similar results for 0.5 s, 1 s and 2 s exposure times. Now only the long exposure time of 5 s differs significantly from the rest. 

For comparison, the mean luminance values in NSU obtained from the luminance maps (Fig. 3), are plotted in Fig. 6 b). Without cosine correction and using all the data, the mean values of the (scalar) illuminance range from 16.4 $\pm$ 1.6 NSU to 21.6 $\pm$ 2.2 NSU, while with cosine correction the values of the (horizontal) illuminance vary only between 6.2 $\pm$ 0.6 NSU and 6.9 ± 0.7 NSU. Two cut-off zenith angles are also shown without cosine correction: only NSB values up to angles of 74$^{\circ}$ and 20$^{\circ}$ from the normal vector have been used. The 20$^{\circ}$ angle is considered because it is similar to the opening angle of the SQM, a commonly used device in NSB measuring networks. The 74$^{\circ}$ was chosen to show up to what angle the NSB in this particular site is relatively homogeneous. The values are summarized in Table 1. In addition to all-sky images, handheld SQM measurements were taken at the same sites. These observations gave values of 6.3 $\pm$ 0.6 NSU (19.6 $\pm$ 0.1 magSQM/arcsec²). This is in good agreement with the values obtained for the 20$^{\circ}$ cut off using the DSLR camera.

\begin{figure}[h]
\centering
a)\includegraphics[width=80mm]{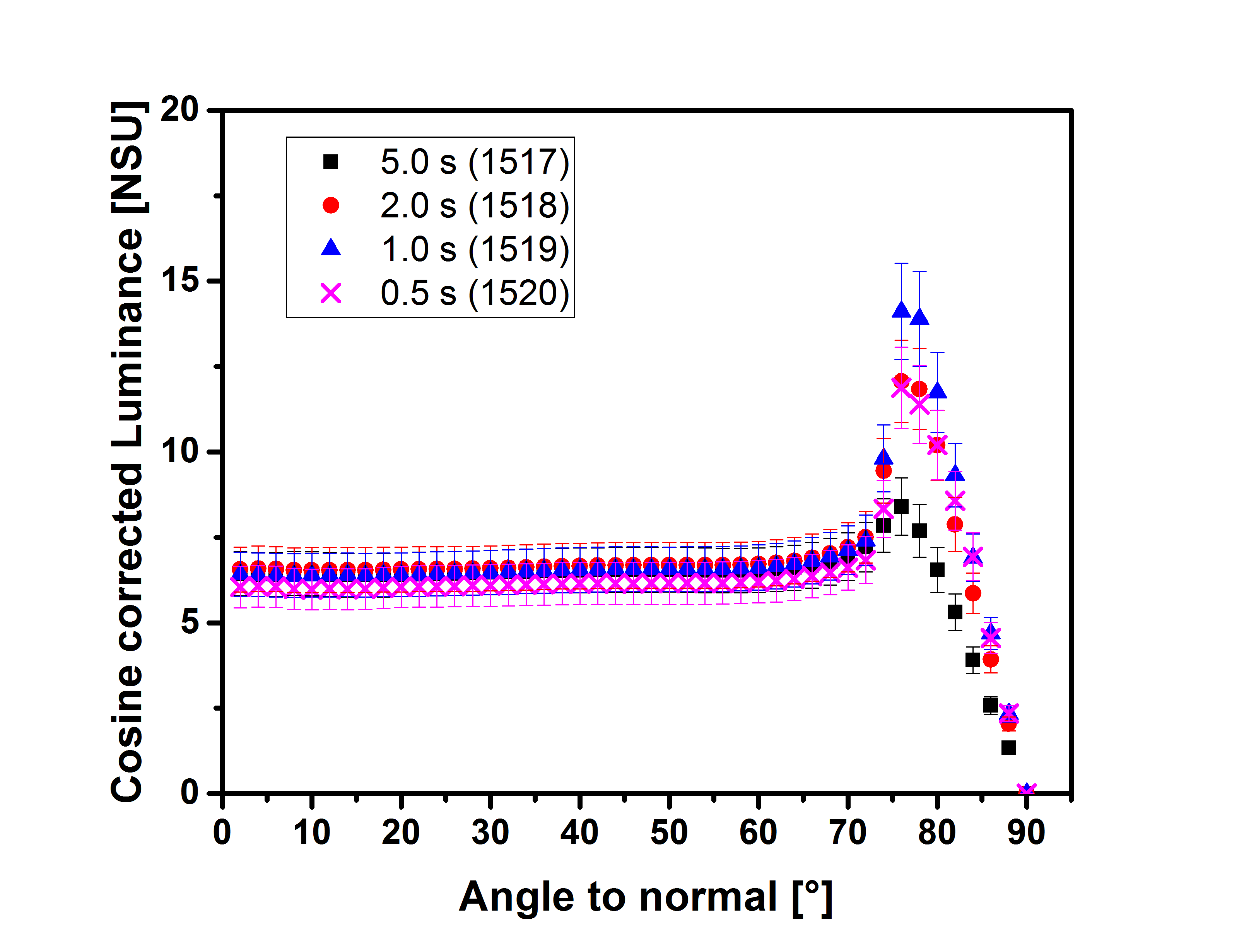}
b)\includegraphics[width=80mm]{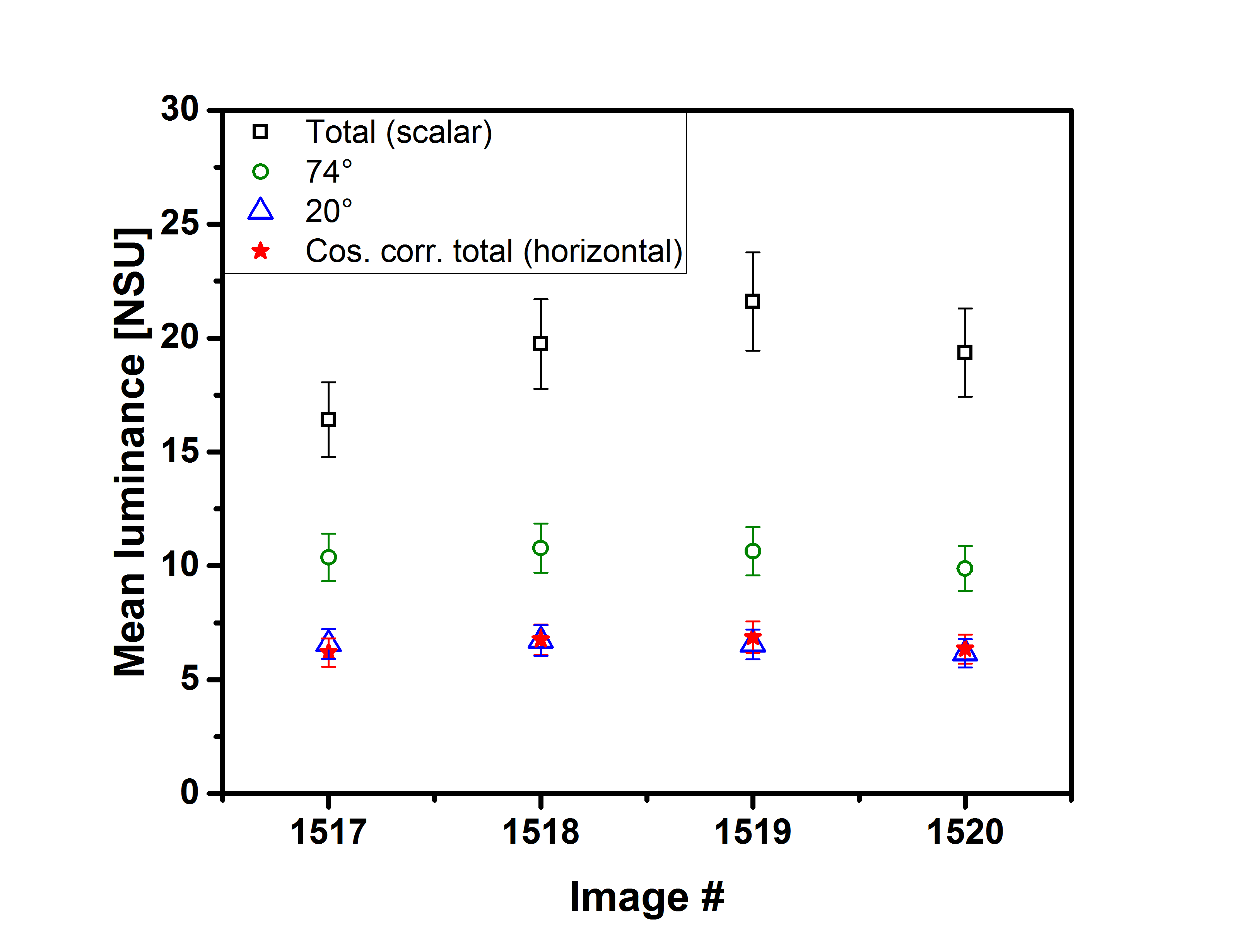}
\caption{Luminance as a function of the angle with respect to the imaging normal vector for, a) the full angular and dynamic range b) zoom in to angles below 70$^{\circ}$,where the values for different exposure times show very good agreement.}
\label{anglu2}
\end{figure}

The error bars and margins in the plots and in the table were chosen to be 10$\%$ due to the error stemming from the calibration procedure. This error and the systematic error from the pointing differences between the images (shaking and rotating boat) is certainly higher than any statistical error in the image analysis. An evaluation of the error between exposure times can only be done with stable illumination conditions in the laboratory. So far we have observed that inter camera values are relatively consistent by comparing many cameras in the field \cite{KybaLonne2015, RibasLonne2017}. Apart from the obvious saturation effects near the horizon especially for image 1517 (5s exposure time), it can be stated that neither shaking of the camera nor change in exposure time does cause an error higher than from the calibration. Considering this, we judge that the spread between the measurements in between the individual images is not very large and that the noise level is acceptable for the short exposure time.
The differences mainly originate from the turning of the boat and from saturation effects. Surprisingly, the relatively short exposure time of 0.5 s at an ISO setting of 6400 can produce reasonable information of the NSB at a site with a luminance of about 6 NSU at zenith and non cosine corrected luminance mean values of 16.4 - 21.6 NSU, while capturing a dynamic range up to 1000 NSU.

\begin{table}[h!]
  \centering
  \caption{List of the locations along the transect from Balaguer to Port d' Àger. Stationary observations were performed during the entire period of the transect from the final location (Parc Astronòmic Montsec, PAM-COU).}
  \label{table_sites}
  \begin{tabular}{cccccc}
    \toprule
   Image $\#$ & Exposure time & &Luminance [NSU]& &\\
	 &  & Total & Cos. corr. & 74$^{\circ}$ cut off & 20$^{\circ}$ cut off \\
   \toprule
1517 (a) & 5 s & 16.4 $\pm$ 1.6	& 6.2 $\pm$ 0.6 &	10.4 $\pm$ 1.0 &	6.6 $\pm$ 0.7\\
1518 (b) & 2 s & 19.7 $\pm$ 2.0	& 6.7 $\pm$ 0.7	& 10.8 $\pm$ 1.1 & 6.7 $\pm$ 0.7\\
1519 (c) & 1 s & 21.6 $\pm$ 2.2	& 6.9 $\pm$ 0.7	& 10.6 $\pm$ 1.1	& 6.6 $\pm$ 0.7\\
1520 (d) & 0.5 s & 19.4 $\pm$ 2.0	& 6.4 $\pm$ 0.6	& 9.9 $\pm$ 1.0	& 6.2 $\pm$ 0.6\\
  \toprule  \end{tabular}
\end{table}

Our data also shows that a zenith measurement (here mimicked with a 20$^{\circ}$ cut off and confirmed with an SQM) has the main drawback that it can underestimate the overall NSB. When inferring the scalar and horizontal illuminance from the zenith NSB, the deviation between the near zenith values and the non cosine corrected integration over the whole hemisphere is up to a factor of three, while the cosine corrected values match for this particular brightness distribution.

\section{Conclusion and outlook}
We have conducted NSB measurements with a commercial DSLR camera from a moving boat. Despite smearing effects, the overall NSB distribution could be obtained from near all-sky brightness maps. This proof of concept study shows that the method is well applicable for investigation of skyglow in marine and also freshwater environments. By undertaking quick measurements from moving vessels it is possible to acquire NSB data that provides spatial information about the LP. By varying the exposure time, it was shown that saturation effects hamper the measurements more than noise. The data demonstrates, that exposure times as low as 0.5 s produce reasonable results, at least for this site and with this camera lens combination, which is commonly used by amateur astronomers. We want to point out, that averaging of many images acquired with short exposure times could improve signal to noise even further, like in the “lucky imaging” technique is used to correct for a turbid and rapidly changing atmosphere \cite{law2006lucky}. 
 While the image quality does not satisfy the standards of conventional astro-photography, this precision is not necessary in the context of ecological LP. For this application, quick data acquisition at many sites and comparability between them is more important than absolute precision. The spatial information is in our opinion needed to infer the propagation of the light and the possible perception of different animals. A simple measurement of the zenith NSB using e.g. an SQM might misinterpret the portion of light at the horizon, and therefore the overall scalar illuminance.
Another potentially interesting application of the method discussed here is the investigation of the propagation of skyglow from single sources into a dark environment. This could be done from an isolated coastal town out into open waters. In contrast to land-based studies, the measurement on the water has the advantage of the absence of obstacles, and the ability to perform a straight transect. For this particular task a more rigid construction, a larger vessel less prone to rocking and drifting, and motion stabilization of the camera are recommended.

\section{Acknowledgements}
This work was supported by ESSEM COST Action ES1204 - Loss of the Night Network (LoNNe). The survey was designed and executed during a COST LoNNe meeting in Israel. AJ, FH and CCMK acknowledge funding by the ILES of the Leibniz Association, Germany (SAW-2015-IGB-1), and the Verlust der Nacht project funded by the Federal Ministry of Education and Research, Germany (BMBF-033L038A).


\begin{thebibliography}{40}
\expandafter\ifx\csname natexlab\endcsname\relax\def\natexlab#1{#1}\fi
\expandafter\ifx\csname bibnamefont\endcsname\relax
  \def\bibnamefont#1{#1}\fi
\expandafter\ifx\csname bibfnamefont\endcsname\relax
  \def\bibfnamefont#1{#1}\fi
\expandafter\ifx\csname citenamefont\endcsname\relax
  \def\citenamefont#1{#1}\fi
\expandafter\ifx\csname url\endcsname\relax
  \def\url#1{\texttt{#1}}\fi
\expandafter\ifx\csname urlprefix\endcsname\relax\def\urlprefix{URL }\fi
\providecommand{\bibinfo}[2]{#2}
\providecommand{\eprint}[2][]{\url{#2}}

\bibitem[{\citenamefont{Stevens and Zhu}(2015)}]{Stevens:2015}
\bibinfo{author}{\bibfnamefont{R.~G.} \bibnamefont{Stevens}} \bibnamefont{and}
  \bibinfo{author}{\bibfnamefont{Y.}~\bibnamefont{Zhu}},
  \bibinfo{journal}{Philosophical Transactions of the Royal Society of London
  B: Biological Sciences} \textbf{\bibinfo{volume}{370}},
  \bibinfo{pages}{20140120} (\bibinfo{year}{2015}).

\bibitem[{\citenamefont{{Narisada} and {Schreuder}}(2004)}]{book:Narisada}
\bibinfo{editor}{\bibfnamefont{K.}~\bibnamefont{{Narisada}}} \bibnamefont{and}
  \bibinfo{editor}{\bibfnamefont{D.}~\bibnamefont{{Schreuder}}}, eds.,
  \emph{\bibinfo{title}{{Light pollution handbook}}}, vol.
  \bibinfo{volume}{322} of \emph{\bibinfo{series}{Astrophysics and Space
  Science Library}} (\bibinfo{publisher}{Springer}, \bibinfo{year}{2004}).

\bibitem[{\citenamefont{H\"olker
  et~al.}(2010{\natexlab{a}})\citenamefont{H\"olker, Moss, Griefahn, Kloas,
  Voigt, Henckel, H\"anel, Kappeler, V\"olker, Schwope
  et~al.}}]{Hoelker:2010_b}
\bibinfo{author}{\bibfnamefont{F.}~\bibnamefont{H\"olker}},
  \bibinfo{author}{\bibfnamefont{T.}~\bibnamefont{Moss}},
  \bibinfo{author}{\bibfnamefont{B.}~\bibnamefont{Griefahn}},
  \bibinfo{author}{\bibfnamefont{W.}~\bibnamefont{Kloas}},
  \bibinfo{author}{\bibfnamefont{C.}~\bibnamefont{Voigt}},
  \bibinfo{author}{\bibfnamefont{A.}~\bibnamefont{Henckel}},
  \bibinfo{author}{\bibfnamefont{A.}~\bibnamefont{H\"anel}},
  \bibinfo{author}{\bibfnamefont{P.}~\bibnamefont{Kappeler}},
  \bibinfo{author}{\bibfnamefont{S.}~\bibnamefont{V\"olker}},
  \bibinfo{author}{\bibfnamefont{A.}~\bibnamefont{Schwope}},
  \bibnamefont{et~al.}, \bibinfo{journal}{Ecology and Society}
  \textbf{\bibinfo{volume}{15}}, \bibinfo{pages}{13}
  (\bibinfo{year}{2010}{\natexlab{a}}).

\bibitem[{\citenamefont{Lyytim\''äki}(20015)}]{Lyyti:2015}
\bibinfo{author}{\bibfnamefont{J.}~\bibnamefont{Lyytim\''äki}},
  \bibinfo{type}{Tech. Rep.}, \bibinfo{institution}{UN-DESA}
  (\bibinfo{year}{20015}).

\bibitem[{\citenamefont{Rich and Longcore}(2006)}]{book:rich_longcore}
\bibinfo{editor}{\bibfnamefont{C.}~\bibnamefont{Rich}} \bibnamefont{and}
  \bibinfo{editor}{\bibfnamefont{T.}~\bibnamefont{Longcore}}, eds.,
  \emph{\bibinfo{title}{Ecological Consequences of Artificial Night Lighting}}
  (\bibinfo{publisher}{Island Press}, \bibinfo{address}{Washington, D.C., USA},
  \bibinfo{year}{2006}).

\bibitem[{\citenamefont{H\"olker
  et~al.}(2010{\natexlab{b}})\citenamefont{H\"olker, Wolter, Perkin, and
  Tockner}}]{Hoelker:2010_a}
\bibinfo{author}{\bibfnamefont{F.}~\bibnamefont{H\"olker}},
  \bibinfo{author}{\bibfnamefont{C.}~\bibnamefont{Wolter}},
  \bibinfo{author}{\bibfnamefont{E.~K.} \bibnamefont{Perkin}},
  \bibnamefont{and} \bibinfo{author}{\bibfnamefont{K.}~\bibnamefont{Tockner}},
  \bibinfo{journal}{Trends in Ecology and Evolution}
  \textbf{\bibinfo{volume}{25}}, \bibinfo{pages}{681}
  (\bibinfo{year}{2010}{\natexlab{b}}).

\bibitem[{\citenamefont{Gaston et~al.}(2015)\citenamefont{Gaston, Visser, and
  H{\"o}lker}}]{Gaston:2015_ptb}
\bibinfo{author}{\bibfnamefont{K.~J.} \bibnamefont{Gaston}},
  \bibinfo{author}{\bibfnamefont{M.~E.} \bibnamefont{Visser}},
  \bibnamefont{and}
  \bibinfo{author}{\bibfnamefont{F.}~\bibnamefont{H{\"o}lker}},
  \bibinfo{journal}{Philosophical Transactions of the Royal Society of London
  B: Biological Sciences} \textbf{\bibinfo{volume}{370}},
  \bibinfo{pages}{20140133} (\bibinfo{year}{2015}).

\bibitem[{\citenamefont{Rowse et~al.}(2016)\citenamefont{Rowse, Lewanzik,
  Stone, Harris, and Jones}}]{rowse2016dark}
\bibinfo{author}{\bibfnamefont{E.}~\bibnamefont{Rowse}},
  \bibinfo{author}{\bibfnamefont{D.}~\bibnamefont{Lewanzik}},
  \bibinfo{author}{\bibfnamefont{E.}~\bibnamefont{Stone}},
  \bibinfo{author}{\bibfnamefont{S.}~\bibnamefont{Harris}}, \bibnamefont{and}
  \bibinfo{author}{\bibfnamefont{G.}~\bibnamefont{Jones}}, in
  \emph{\bibinfo{booktitle}{Bats in the Anthropocene: Conservation of Bats in a
  Changing World}} (\bibinfo{publisher}{Springer}, \bibinfo{year}{2016}), pp.
  \bibinfo{pages}{187--213}.

\bibitem[{\citenamefont{Robert et~al.}(2015)\citenamefont{Robert, Lesku,
  Partecke, and Chambers}}]{robert2015artificial}
\bibinfo{author}{\bibfnamefont{K.~A.} \bibnamefont{Robert}},
  \bibinfo{author}{\bibfnamefont{J.~A.} \bibnamefont{Lesku}},
  \bibinfo{author}{\bibfnamefont{J.}~\bibnamefont{Partecke}}, \bibnamefont{and}
  \bibinfo{author}{\bibfnamefont{B.}~\bibnamefont{Chambers}}, in
  \emph{\bibinfo{booktitle}{Proc. R. Soc. B}} (\bibinfo{organization}{The Royal
  Society}, \bibinfo{year}{2015}), vol. \bibinfo{volume}{282}, p.
  \bibinfo{pages}{20151745}.

\bibitem[{\citenamefont{Macgregor et~al.}(2015)\citenamefont{Macgregor, Pocock,
  Fox, and Evans}}]{macgregor2015pollination}
\bibinfo{author}{\bibfnamefont{C.~J.} \bibnamefont{Macgregor}},
  \bibinfo{author}{\bibfnamefont{M.~J.} \bibnamefont{Pocock}},
  \bibinfo{author}{\bibfnamefont{R.}~\bibnamefont{Fox}}, \bibnamefont{and}
  \bibinfo{author}{\bibfnamefont{D.~M.} \bibnamefont{Evans}},
  \bibinfo{journal}{Ecological entomology} \textbf{\bibinfo{volume}{40}},
  \bibinfo{pages}{187} (\bibinfo{year}{2015}).

\bibitem[{\citenamefont{Bennie et~al.}(2016)\citenamefont{Bennie, Davies,
  Cruse, and Gaston}}]{bennie2016plants}
\bibinfo{author}{\bibfnamefont{J.}~\bibnamefont{Bennie}},
  \bibinfo{author}{\bibfnamefont{T.~W.} \bibnamefont{Davies}},
  \bibinfo{author}{\bibfnamefont{D.}~\bibnamefont{Cruse}}, \bibnamefont{and}
  \bibinfo{author}{\bibfnamefont{K.~J.} \bibnamefont{Gaston}},
  \bibinfo{journal}{Journal of Ecology}  (\bibinfo{year}{2016}).

\bibitem[{\citenamefont{Davies et~al.}(2014)\citenamefont{Davies, Duffy,
  Bennie, and Gaston}}]{davies2014marine}
\bibinfo{author}{\bibfnamefont{T.~W.} \bibnamefont{Davies}},
  \bibinfo{author}{\bibfnamefont{J.~P.} \bibnamefont{Duffy}},
  \bibinfo{author}{\bibfnamefont{J.}~\bibnamefont{Bennie}}, \bibnamefont{and}
  \bibinfo{author}{\bibfnamefont{K.~J.} \bibnamefont{Gaston}},
  \bibinfo{journal}{Frontiers in Ecology and the Environment}
  \textbf{\bibinfo{volume}{12}}, \bibinfo{pages}{347} (\bibinfo{year}{2014}).

\bibitem[{\citenamefont{Perkin et~al.}(2011)\citenamefont{Perkin, H\"olker,
  Richardson, Sadler, Wolter, and Tockner}}]{Perkin:2011}
\bibinfo{author}{\bibfnamefont{E.}~\bibnamefont{Perkin}},
  \bibinfo{author}{\bibfnamefont{F.}~\bibnamefont{H\"olker}},
  \bibinfo{author}{\bibfnamefont{J.}~\bibnamefont{Richardson}},
  \bibinfo{author}{\bibfnamefont{J.}~\bibnamefont{Sadler}},
  \bibinfo{author}{\bibfnamefont{C.}~\bibnamefont{Wolter}}, \bibnamefont{and}
  \bibinfo{author}{\bibfnamefont{K.}~\bibnamefont{Tockner}},
  \bibinfo{journal}{Ecosphere} \textbf{\bibinfo{volume}{2}},
  \bibinfo{pages}{122} (\bibinfo{year}{2011}).

\bibitem[{\citenamefont{Moore et~al.}(2000)\citenamefont{Moore, Pierce, Walsh,
  Kvalvik, and Lim}}]{Moore:2000}
\bibinfo{author}{\bibfnamefont{M.~V.} \bibnamefont{Moore}},
  \bibinfo{author}{\bibfnamefont{S.~M.} \bibnamefont{Pierce}},
  \bibinfo{author}{\bibfnamefont{H.~M.} \bibnamefont{Walsh}},
  \bibinfo{author}{\bibfnamefont{S.~K.} \bibnamefont{Kvalvik}},
  \bibnamefont{and} \bibinfo{author}{\bibfnamefont{J.~D.} \bibnamefont{Lim}},
  \bibinfo{journal}{Verhandlungen der Internatinoalen Vereinigung f\"ur
  Theoretische und Angewandte Limnologie} \textbf{\bibinfo{volume}{27}},
  \bibinfo{pages}{779} (\bibinfo{year}{2000}).

\bibitem[{\citenamefont{Bruening et~al.}(2015)\citenamefont{Bruening, Hoelker,
  Franke, Preuer, and Kloas}}]{bruening2015spotlight}
\bibinfo{author}{\bibfnamefont{A.}~\bibnamefont{Bruening}},
  \bibinfo{author}{\bibfnamefont{F.}~\bibnamefont{Hoelker}},
  \bibinfo{author}{\bibfnamefont{S.}~\bibnamefont{Franke}},
  \bibinfo{author}{\bibfnamefont{T.}~\bibnamefont{Preuer}}, \bibnamefont{and}
  \bibinfo{author}{\bibfnamefont{W.}~\bibnamefont{Kloas}},
  \bibinfo{journal}{Science of the Total Environment}
  \textbf{\bibinfo{volume}{511}}, \bibinfo{pages}{516} (\bibinfo{year}{2015}).

\bibitem[{\citenamefont{Jechow et~al.}(2016{\natexlab{a}})\citenamefont{Jechow,
  Hölker, Kolláth, Gessner, and Kyba}}]{Jechow2016}
\bibinfo{author}{\bibfnamefont{A.}~\bibnamefont{Jechow}},
  \bibinfo{author}{\bibfnamefont{F.}~\bibnamefont{Hölker}},
  \bibinfo{author}{\bibfnamefont{Z.}~\bibnamefont{Kolláth}},
  \bibinfo{author}{\bibfnamefont{M.~O.} \bibnamefont{Gessner}},
  \bibnamefont{and} \bibinfo{author}{\bibfnamefont{C.~C.~M.}
  \bibnamefont{Kyba}}, \bibinfo{journal}{Journal of Quantitative Spectroscopy
  and Radiative Transfer} \textbf{\bibinfo{volume}{181}}, \bibinfo{pages}{24 }
  (\bibinfo{year}{2016}{\natexlab{a}}).

\bibitem[{\citenamefont{Riegel}(1973)}]{Riegel1973}
\bibinfo{author}{\bibfnamefont{K.~W.} \bibnamefont{Riegel}},
  \bibinfo{journal}{Science} \textbf{\bibinfo{volume}{179}},
  \bibinfo{pages}{1285} (\bibinfo{year}{1973}).

\bibitem[{\citenamefont{Berry}(1976)}]{berry1976light}
\bibinfo{author}{\bibfnamefont{R.~L.} \bibnamefont{Berry}},
  \bibinfo{journal}{Journal of the Royal Astronomical Society of Canada}
  \textbf{\bibinfo{volume}{70}}, \bibinfo{pages}{97} (\bibinfo{year}{1976}).

\bibitem[{\citenamefont{Kyba et~al.}(2011)\citenamefont{Kyba, Ruhtz, Fischer,
  and H\"olker}}]{Kyba:2011_sqm}
\bibinfo{author}{\bibfnamefont{C.~C.~M.} \bibnamefont{Kyba}},
  \bibinfo{author}{\bibfnamefont{T.}~\bibnamefont{Ruhtz}},
  \bibinfo{author}{\bibfnamefont{J.}~\bibnamefont{Fischer}}, \bibnamefont{and}
  \bibinfo{author}{\bibfnamefont{F.}~\bibnamefont{H\"olker}},
  \bibinfo{journal}{PLOS ONE} \textbf{\bibinfo{volume}{6}},
  \bibinfo{pages}{e17307} (\bibinfo{year}{2011}).

\bibitem[{\citenamefont{Kocifaj and Lamphar}(2014)}]{Kocifaj:2014_cloud_theory}
\bibinfo{author}{\bibfnamefont{M.}~\bibnamefont{Kocifaj}} \bibnamefont{and}
  \bibinfo{author}{\bibfnamefont{H.~A.~S.} \bibnamefont{Lamphar}},
  \bibinfo{journal}{Monthly Notices of the Royal Astronomical Society}
  \textbf{\bibinfo{volume}{443}}, \bibinfo{pages}{3665} (\bibinfo{year}{2014}).

\bibitem[{\citenamefont{Jechow et~al.}(2016{\natexlab{b}})\citenamefont{Jechow,
  Kolláth, Hölker, and Kyba}}]{jechowALAN}
\bibinfo{author}{\bibfnamefont{A.}~\bibnamefont{Jechow}},
  \bibinfo{author}{\bibfnamefont{Z.}~\bibnamefont{Kolláth}},
  \bibinfo{author}{\bibfnamefont{F.}~\bibnamefont{Hölker}}, \bibnamefont{and}
  \bibinfo{author}{\bibfnamefont{C.~C.~M.} \bibnamefont{Kyba}}, in
  \emph{\bibinfo{booktitle}{ALAN Conference 2016}}
  (\bibinfo{year}{2016}{\natexlab{b}}).

\bibitem[{\citenamefont{Ribas et~al.}(2016)\citenamefont{Ribas, Figueras,
  Paricio, Canal-Domingo, and Torra}}]{Ribas2016clouds}
\bibinfo{author}{\bibfnamefont{S.~J.} \bibnamefont{Ribas}},
  \bibinfo{author}{\bibfnamefont{F.}~\bibnamefont{Figueras}},
  \bibinfo{author}{\bibfnamefont{S.}~\bibnamefont{Paricio}},
  \bibinfo{author}{\bibfnamefont{R.}~\bibnamefont{Canal-Domingo}},
  \bibnamefont{and} \bibinfo{author}{\bibfnamefont{J.}~\bibnamefont{Torra}},
  \bibinfo{journal}{International Journal of Sustainable Lighting}
  \textbf{\bibinfo{volume}{35}}, \bibinfo{pages}{32} (\bibinfo{year}{2016}).

\bibitem[{\citenamefont{Falchi et~al.}(2016)\citenamefont{Falchi, Cinzano,
  Duriscoe, Kyba, Elvidge, Baugh, Portnov, Rybnikova, and
  Furgoni}}]{falchi2016WA}
\bibinfo{author}{\bibfnamefont{F.}~\bibnamefont{Falchi}},
  \bibinfo{author}{\bibfnamefont{P.}~\bibnamefont{Cinzano}},
  \bibinfo{author}{\bibfnamefont{D.}~\bibnamefont{Duriscoe}},
  \bibinfo{author}{\bibfnamefont{C.~C.} \bibnamefont{Kyba}},
  \bibinfo{author}{\bibfnamefont{C.~D.} \bibnamefont{Elvidge}},
  \bibinfo{author}{\bibfnamefont{K.}~\bibnamefont{Baugh}},
  \bibinfo{author}{\bibfnamefont{B.~A.} \bibnamefont{Portnov}},
  \bibinfo{author}{\bibfnamefont{N.~A.} \bibnamefont{Rybnikova}},
  \bibnamefont{and} \bibinfo{author}{\bibfnamefont{R.}~\bibnamefont{Furgoni}},
  \bibinfo{journal}{Science advances} \textbf{\bibinfo{volume}{2}},
  \bibinfo{pages}{e1600377} (\bibinfo{year}{2016}).

\bibitem[{\citenamefont{Cinzano}(2005)}]{Cinzano_sqm:2005}
\bibinfo{author}{\bibfnamefont{P.}~\bibnamefont{Cinzano}}, \bibinfo{type}{Tech.
  Rep.} \bibinfo{number}{9}, \bibinfo{institution}{ISTIL}
  (\bibinfo{year}{2005}), \bibinfo{note}{v1.4}.

\bibitem[{\citenamefont{Kyba et~al.}(2015{\natexlab{a}})\citenamefont{Kyba,
  Tong, Bennie, Birriel, Birriel, Cool, Danielsen, Davies, Peter, Edwards
  et~al.}}]{Kyba:2015_isqm}
\bibinfo{author}{\bibfnamefont{C.~C.~M.} \bibnamefont{Kyba}},
  \bibinfo{author}{\bibfnamefont{K.~P.} \bibnamefont{Tong}},
  \bibinfo{author}{\bibfnamefont{J.}~\bibnamefont{Bennie}},
  \bibinfo{author}{\bibfnamefont{I.}~\bibnamefont{Birriel}},
  \bibinfo{author}{\bibfnamefont{J.~J.} \bibnamefont{Birriel}},
  \bibinfo{author}{\bibfnamefont{A.}~\bibnamefont{Cool}},
  \bibinfo{author}{\bibfnamefont{A.}~\bibnamefont{Danielsen}},
  \bibinfo{author}{\bibfnamefont{T.~W.} \bibnamefont{Davies}},
  \bibinfo{author}{\bibfnamefont{N.}~\bibnamefont{Peter}},
  \bibinfo{author}{\bibfnamefont{W.}~\bibnamefont{Edwards}},
  \bibnamefont{et~al.}, \bibinfo{journal}{Scientific Reports}
  \textbf{\bibinfo{volume}{5}}, \bibinfo{pages}{8409}
  (\bibinfo{year}{2015}{\natexlab{a}}).

\bibitem[{\citenamefont{Pun et~al.}(2014)\citenamefont{Pun, So, Leung, and
  Wong}}]{Pun:2014}
\bibinfo{author}{\bibfnamefont{C.}~\bibnamefont{Pun}},
  \bibinfo{author}{\bibfnamefont{C.}~\bibnamefont{So}},
  \bibinfo{author}{\bibfnamefont{W.}~\bibnamefont{Leung}}, \bibnamefont{and}
  \bibinfo{author}{\bibfnamefont{C.}~\bibnamefont{Wong}},
  \bibinfo{journal}{Journal of Quantitative Spectroscopy and Radiative
  Transfer} \textbf{\bibinfo{volume}{139}}, \bibinfo{pages}{90}
  (\bibinfo{year}{2014}).

\bibitem[{\citenamefont{Kocifaj et~al.}(2015)\citenamefont{Kocifaj, Posch, and
  Lamphar}}]{kocifaj2015zenith}
\bibinfo{author}{\bibfnamefont{M.}~\bibnamefont{Kocifaj}},
  \bibinfo{author}{\bibfnamefont{T.}~\bibnamefont{Posch}}, \bibnamefont{and}
  \bibinfo{author}{\bibfnamefont{H.~S.} \bibnamefont{Lamphar}},
  \bibinfo{journal}{Monthly Notices of the Royal Astronomical Society}
  \textbf{\bibinfo{volume}{446}}, \bibinfo{pages}{2895} (\bibinfo{year}{2015}).

\bibitem[{\citenamefont{Koll\'{a}th}(2010)}]{Kollath:2010}
\bibinfo{author}{\bibfnamefont{Z.}~\bibnamefont{Koll\'{a}th}},
  \bibinfo{journal}{Journal of Physics: Conference Series}
  \textbf{\bibinfo{volume}{218}}, \bibinfo{pages}{012001}
  (\bibinfo{year}{2010}).

\bibitem[{\citenamefont{{Duriscoe} et~al.}(2007)\citenamefont{{Duriscoe},
  {Luginbuhl}, and {Moore}}}]{Duriscoe:2007}
\bibinfo{author}{\bibfnamefont{D.~M.} \bibnamefont{{Duriscoe}}},
  \bibinfo{author}{\bibfnamefont{C.~B.} \bibnamefont{{Luginbuhl}}},
  \bibnamefont{and} \bibinfo{author}{\bibfnamefont{C.~A.}
  \bibnamefont{{Moore}}}, \bibinfo{journal}{Publications of the Astronomical
  Society of the Pacific} \textbf{\bibinfo{volume}{119}}, \bibinfo{pages}{192}
  (\bibinfo{year}{2007}).

\bibitem[{\citenamefont{Romanishin}(2006)}]{romanishin2006introduction}
\bibinfo{author}{\bibfnamefont{W.}~\bibnamefont{Romanishin}}
  (\bibinfo{year}{2006}).

\bibitem[{\citenamefont{Aceituno et~al.}(2011)\citenamefont{Aceituno,
  S{\'a}nchez, Aceituno, Galad{\'\i}-Enr{\'\i}quez, Negro, Soriguer, and
  Gomez}}]{aceituno2011all}
\bibinfo{author}{\bibfnamefont{J.}~\bibnamefont{Aceituno}},
  \bibinfo{author}{\bibfnamefont{S.}~\bibnamefont{S{\'a}nchez}},
  \bibinfo{author}{\bibfnamefont{F.~J.} \bibnamefont{Aceituno}},
  \bibinfo{author}{\bibfnamefont{D.}~\bibnamefont{Galad{\'\i}-Enr{\'\i}quez}},
  \bibinfo{author}{\bibfnamefont{J.~J.} \bibnamefont{Negro}},
  \bibinfo{author}{\bibfnamefont{R.~C.} \bibnamefont{Soriguer}},
  \bibnamefont{and} \bibinfo{author}{\bibfnamefont{G.~S.} \bibnamefont{Gomez}},
  \bibinfo{journal}{Publications of the Astronomical Society of the Pacific}
  \textbf{\bibinfo{volume}{123}}, \bibinfo{pages}{1076} (\bibinfo{year}{2011}).

\bibitem[{\citenamefont{Aub{\'e} et~al.}(2016)\citenamefont{Aub{\'e}, Kocifaj,
  Zamorano, Solano-Lamphar, and de~Miguel}}]{aube2016spectral}
\bibinfo{author}{\bibfnamefont{M.}~\bibnamefont{Aub{\'e}}},
  \bibinfo{author}{\bibfnamefont{M.}~\bibnamefont{Kocifaj}},
  \bibinfo{author}{\bibfnamefont{J.}~\bibnamefont{Zamorano}},
  \bibinfo{author}{\bibfnamefont{H.}~\bibnamefont{Solano-Lamphar}},
  \bibnamefont{and} \bibinfo{author}{\bibfnamefont{A.~S.}
  \bibnamefont{de~Miguel}}, \bibinfo{journal}{Journal of Quantitative
  Spectroscopy and Radiative Transfer} \textbf{\bibinfo{volume}{181}},
  \bibinfo{pages}{11} (\bibinfo{year}{2016}).

\bibitem[{\citenamefont{Kyba et~al.}(2015{\natexlab{b}})\citenamefont{Kyba,
  Bouroussis, Canal-Domingo, Falchi, Giacomelli, H{\"a}nel, Koll{\'a}th,
  Massetti, Ribas, Spoelstra et~al.}}]{KybaLonne2015}
\bibinfo{author}{\bibfnamefont{C.}~\bibnamefont{Kyba}},
  \bibinfo{author}{\bibfnamefont{C.}~\bibnamefont{Bouroussis}},
  \bibinfo{author}{\bibfnamefont{R.}~\bibnamefont{Canal-Domingo}},
  \bibinfo{author}{\bibfnamefont{F.}~\bibnamefont{Falchi}},
  \bibinfo{author}{\bibfnamefont{A.}~\bibnamefont{Giacomelli}},
  \bibinfo{author}{\bibfnamefont{A.}~\bibnamefont{H{\"a}nel}},
  \bibinfo{author}{\bibfnamefont{Z.}~\bibnamefont{Koll{\'a}th}},
  \bibinfo{author}{\bibfnamefont{L.}~\bibnamefont{Massetti}},
  \bibinfo{author}{\bibfnamefont{S.~J.} \bibnamefont{Ribas}},
  \bibinfo{author}{\bibfnamefont{H.}~\bibnamefont{Spoelstra}},
  \bibnamefont{et~al.}, \bibinfo{type}{Tech. Rep.},
  \bibinfo{institution}{LoNNe}, \bibinfo{address}{GFZ Potsdam}
  (\bibinfo{year}{2015}{\natexlab{b}}).

\bibitem[{\citenamefont{Ribas et~al.}(2017)\citenamefont{Ribas, Aub{\'e},
  Bar{\'a}, Bouroussis, Canal-Domingo, Espey, H{\"a}nel, Jechow, Koll{\'a}th,
  Marti et~al.}}]{RibasLonne2017}
\bibinfo{author}{\bibfnamefont{S.~J.} \bibnamefont{Ribas}},
  \bibinfo{author}{\bibfnamefont{M.}~\bibnamefont{Aub{\'e}}},
  \bibinfo{author}{\bibfnamefont{S.}~\bibnamefont{Bar{\'a}}},
  \bibinfo{author}{\bibfnamefont{C.}~\bibnamefont{Bouroussis}},
  \bibinfo{author}{\bibfnamefont{R.}~\bibnamefont{Canal-Domingo}},
  \bibinfo{author}{\bibfnamefont{B.}~\bibnamefont{Espey}},
  \bibinfo{author}{\bibfnamefont{A.}~\bibnamefont{H{\"a}nel}},
  \bibinfo{author}{\bibfnamefont{A.}~\bibnamefont{Jechow}},
  \bibinfo{author}{\bibfnamefont{Z.}~\bibnamefont{Koll{\'a}th}},
  \bibinfo{author}{\bibfnamefont{G.}~\bibnamefont{Marti}},
  \bibnamefont{et~al.}, \bibinfo{type}{Tech. Rep.},
  \bibinfo{institution}{LoNNe}, \bibinfo{address}{GFZ Potsdam}
  (\bibinfo{year}{2017}).

\bibitem[{\citenamefont{Pogson}(1856)}]{pogson1856magnitudes}
\bibinfo{author}{\bibfnamefont{N.}~\bibnamefont{Pogson}},
  \bibinfo{journal}{Monthly Notices of the Royal Astronomical Society}
  \textbf{\bibinfo{volume}{17}}, \bibinfo{pages}{12} (\bibinfo{year}{1856}).

\bibitem[{\citenamefont{H\''anel}(2017 (in submission))}]{haenel2017}
\bibinfo{author}{\bibnamefont{H\''anel}}, \bibinfo{journal}{Journal of
  Quantitative Spectroscopy and Radiative Transfer}  (\bibinfo{year}{2017 (in
  submission)}).

\bibitem[{\citenamefont{Koll{\'a}th}(2017 (in submission))}]{kollath2017}
\bibinfo{author}{\bibfnamefont{Z.}~\bibnamefont{Koll{\'a}th}},
  \bibinfo{journal}{International Journal of Sustainable Lighting}
  (\bibinfo{year}{2017 (in submission)}).

\bibitem[{\citenamefont{Eaton}(1997)}]{eaton1997gnu}
\bibinfo{author}{\bibfnamefont{J.~W.} \bibnamefont{Eaton}},
  \emph{\bibinfo{title}{Gnu octave}} (\bibinfo{year}{1997}).

\bibitem[{\citenamefont{Duriscoe}(2016)}]{duriscoe2016photometric}
\bibinfo{author}{\bibfnamefont{D.~M.} \bibnamefont{Duriscoe}},
  \bibinfo{journal}{Journal of Quantitative Spectroscopy and Radiative
  Transfer} \textbf{\bibinfo{volume}{181}}, \bibinfo{pages}{33}
  (\bibinfo{year}{2016}).

\bibitem[{\citenamefont{Law et~al.}(2006)\citenamefont{Law, Mackay, and
  Baldwin}}]{law2006lucky}
\bibinfo{author}{\bibfnamefont{N.~M.} \bibnamefont{Law}},
  \bibinfo{author}{\bibfnamefont{C.~D.} \bibnamefont{Mackay}},
  \bibnamefont{and} \bibinfo{author}{\bibfnamefont{J.~E.}
  \bibnamefont{Baldwin}}, \bibinfo{journal}{Astronomy \& Astrophysics}
  \textbf{\bibinfo{volume}{446}}, \bibinfo{pages}{739} (\bibinfo{year}{2006}).

\end{thebibliography}
\end{document}